\documentclass[twocolumn]{svjour3}         
\smartqed  
\usepackage{graphicx}
\begin{document}

\title{Are truncated stellar disks linked to the molecular gas density?
}

\titlerunning{Truncated disks}        

\author{V. Casasola,        
        F. Combes,
        D. Bettoni, 
        M. Pohlen,
        G. Galletta \\ and
        F. Tenaglia
}

\authorrunning{Casasola et al.} 

\institute{Viviana Casasola \at
              Padua University, Dep. of Astronomy\\
              Vicolo dell'Osservatorio 2, I-35122 Padua \\
              Tel.: +39-049-8278241\\
              Fax: +39-049-8278212\\
              \email{viviana.casasola@unipd.}           \\
             \emph{Present address: Observatoire de Paris-LERMA\\
              61 Av. de l'Observatoire\\
              F-75014 Paris}   
}

\date{Received: date / Accepted: date}

\maketitle

\begin{abstract}
We know that the slope of the radial, stellar light distribution in galaxies 
is well described by an exponential decline and this distribution is often 
truncated at a break radius 
($R_{br}$). We don'\-t  have a clear understanding for the origin of these outer truncations 
and several hypotheses have been proposed to explain them. We want to test the various theories with 
direct observations of the cold molecular gas for a few truncated galaxies in 
comparison with the non-truncated ones. The answer to the existence of a possible 
link between truncated stellar disks and the molecular gas density cannot be obtained 
from CO maps in the literature, because so far there are no galaxies with a clear truncation 
observed in CO at high resolution.

\end{abstract}

\section{Introduction}
\label{intro}
The properties of the faintest regions of galactic disks 
are linked to the mechanisms involved in the growing and shaping
of galaxies. These outer parts are affected by interactions
with the nearby galaxies and so their characteristics are connected
with the evolution followed by the galaxies. In other words, the stellar 
structure of the outer parts of galactic disks is a fundamental element 
of evidence for understanding the formation and evolution of spiral galaxies. 
Since Patterson (1940), de Vaucouleurs (1959) and Freeman (1970) we 
know that the slope of the radial, stellar light distribution is described by an 
exponential decline. However, this exponential light distribution 
does not always extend to arbitrarily large radii, but it is often 
truncated at a well-defined $R_{br}$ (van der Kruit 1979, de Grijs et al.~2001; Pohlen et 
al.~2002). In the optical this break occurs at galactocentric distances 
of typically 2-3 radial scale-lengths (Pohlen \& Trujillo 2006) and it is 
best described with a broken exponential with a clear break and a 
{\it downbending}, steeper outer region (see Fig. 1). This truncation radius could be 
used as a new intrinsic parameter measuring the size of galactic disks and may be 
linked to their formation and evolution as shown by 
Trujillo \& Pohlen (2005). At present, we do not have a secure understanding for the 
origin of these {\it outer edges}, and the hypotheses proposed range from explaining 
truncations as remains of the initial galaxy formation process (van der Kruit 1987) being
a relic of the galaxy evolution over the lifetime (Seiden et al. 1984).
Another possible and interesting explanation assumes the existence of a critical CO 
density acting as a star formation threshold. If the observed falloff in the star 
formation rate (SFR) in the outer parts persists for sufficient time, it should 
introduce a visible truncation of the stellar luminosity profile at that radius. 
A typical threshold of the molecular gas density, below which the star formation process
is inhibited, is in the range from 3 to 10 M$_\odot$ pc$^{-2}$ according to Schaye (2004). 
This threshold may be observed by measuring the local column density of molecular 
gas and correlating it with the current SFR in the same galaxy regions. 
We are interested to find a change in the local CO emission across 
the optical break radius (or the typical position for it), while comparing 
truncated with non-truncated galaxies. 

\begin{figure*}[ht]
\centering
\begin{tabular}{l l}
  \includegraphics[height=0.48\textwidth,angle=-90]{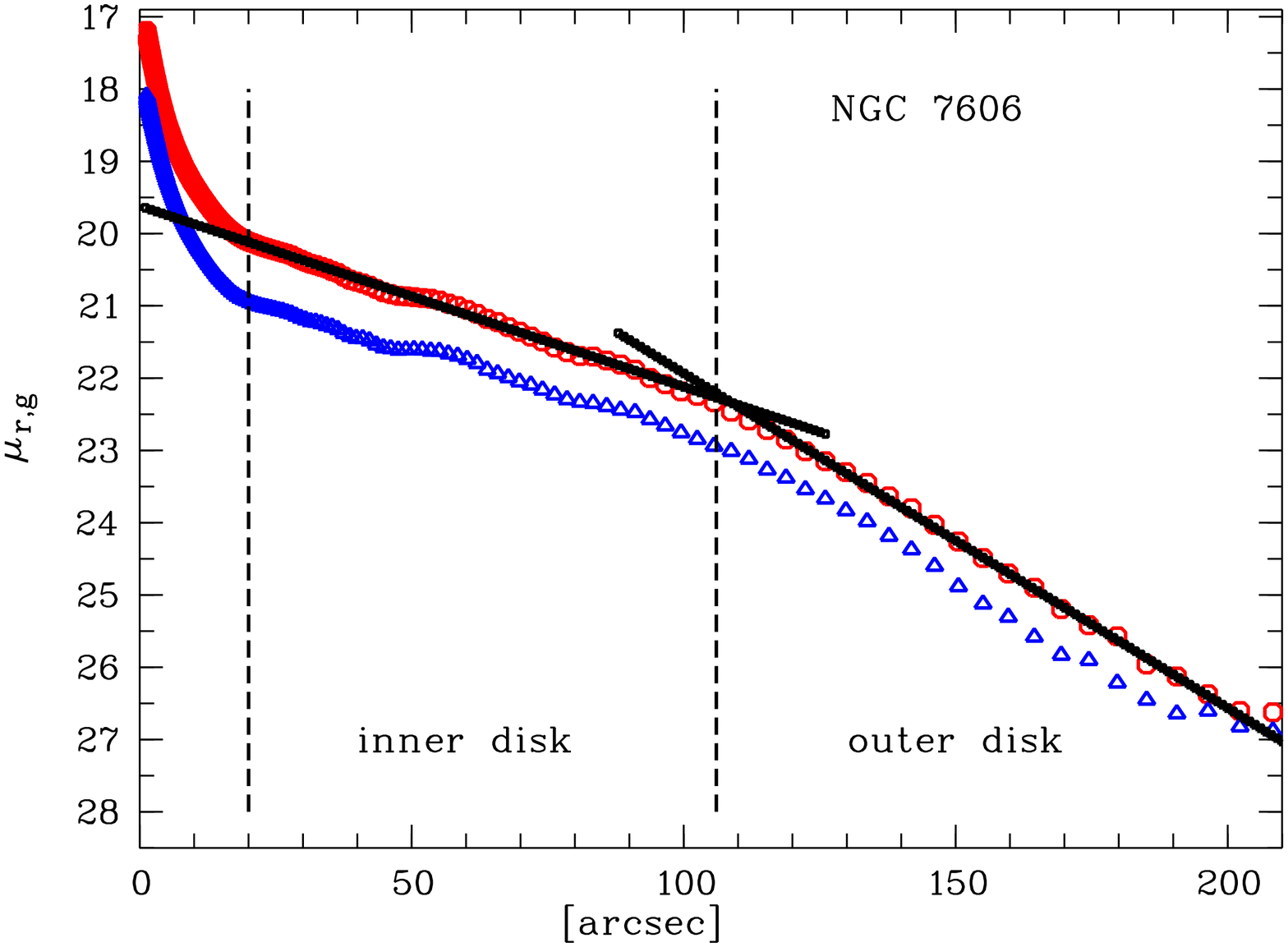}&
 \includegraphics[height=0.48\textwidth,angle=-90]{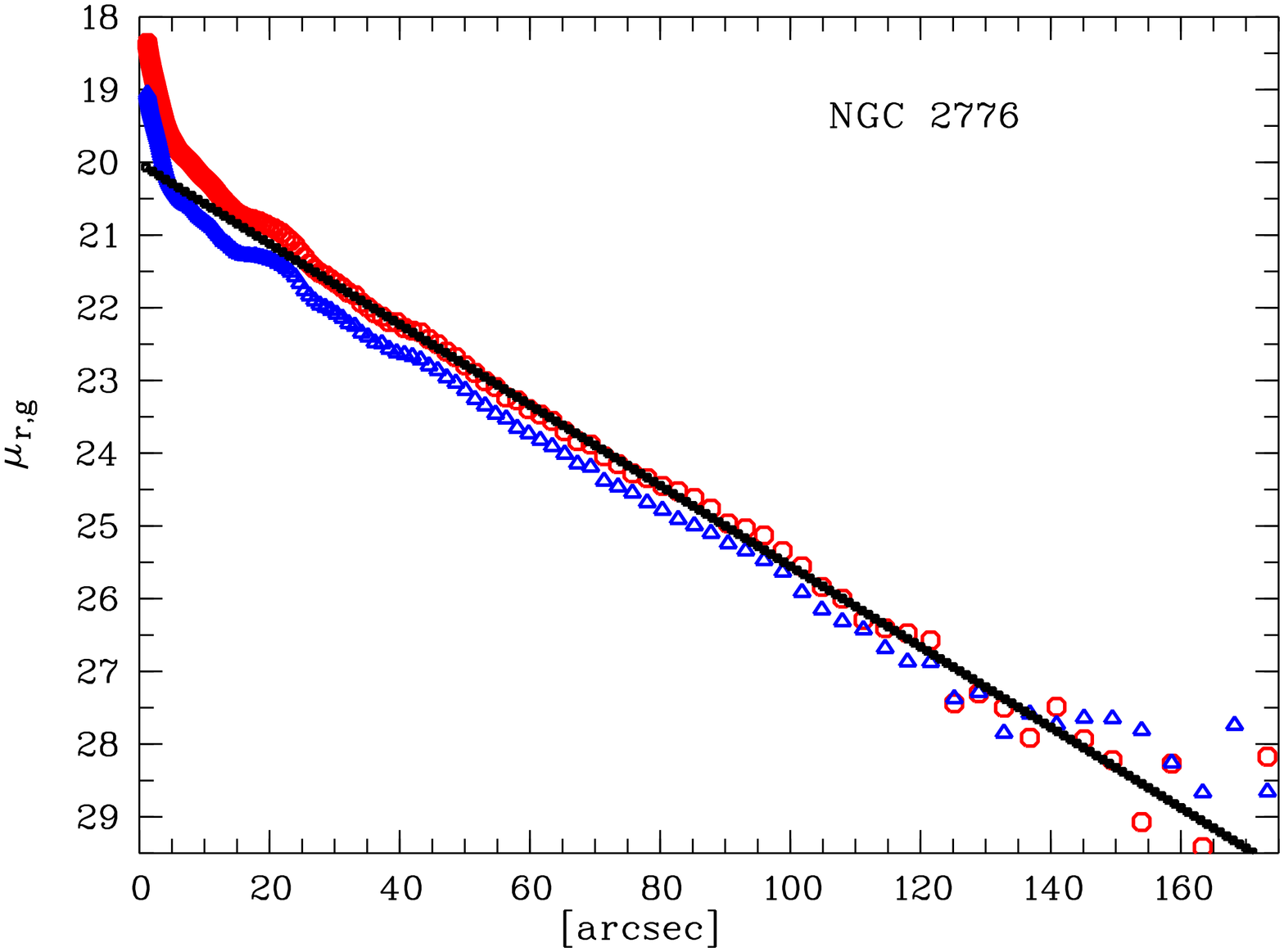}
\end{tabular}
\caption{Azimuthally averaged radial, SLOAN g$^{\prime}$ (blue triangles) and 
r$^{\prime}$ (red circles) -band, surface brightness profiles for NGC7606 and NGC2776.
For the truncated galaxy NGC7606 (left panel) beyond the inner region 
($<20^{\prime\prime}$), dominated by the bulge component, there is an 
inner exponential distribution {\it (inner disk)} out to a break radius at
$R_{\rm br}\!=\!106^{\prime\prime}$ followed by another outer steeper exponential
region {\it (outer disk)}. The solid black lines are exponential fits 
to the associated radial regions.}
\label{fig:1}       
\end{figure*}

\section{Sample Selection}
\label{sec:1}
We selected two galaxies, one with truncation and one without truncation,
from a sample of 98 nearby disk galaxies (Pohlen \& Trujillo 2006) 
with measured deep optical profiles whose inclination is ranging 
from intermediate ($i=70^\circ$) to almost face-on ($i\sim10^\circ$):
NGC7606 (truncated) and NGC2776 (non-truncated). These two galaxies also have 
the highest possible FIR luminosity and H$_2$ gas content in the literature 
and this provides the highest possible chances to find enough CO in these 
galaxies to get a good measure along the radial axis. In Fig. 1 the azimuthally 
averaged radial surface brightness profiles using SDSS data are represented for 
the two galaxies. Comparing the profiles of the two galaxies, one  
clearly see that for the truncated galaxy NGC7606 (left panel) beyond the inner 
region ($<20^{\prime\prime}$), dominated by the bulge component, there is an inner exponential 
distribution {\it (inner disk)} out to a break radius at $R_{\rm br}\!=\!106^{\prime\prime}$ 
followed by another outer steeper exponential region {\it (outer disk)}. 

\section{Observations and First Results}
\label{sec:2}
To test the possible local correlation between CO and SFR, we have
started to observe the two galaxies selected at CO(1-0) and CO(2-1) lines with the 
30m-IRAM Telescope. In particular, we observed the nucleus and some regions 
near the major axis of the two galaxies. For NGC7606 the observational data now available 
(Fig. 2, upper panels) suggest the 
presence of a CO ring between 50$^{\prime\prime}$ and 90$^{\prime\prime}$ 
from the nucleus and the absence of signal 
for larger distances (similar to Fig.1, left panel). NGC2776 shows a CO peak in the 
nucleus and the molecular gas content decreases with radius (Fig. 2, bottom panels) 
with a behavior similar to that obtained from SDSS data (Fig. 1, right panel) in 
g$^{\prime}$ (blue-green) and r$^{\prime}$ (red) wavebands. 

\section{Discussion}
\label{sec:3}
The fact that the truncated galaxy NGC7606 presents only a marginal or none 
CO detection in the center is not so anomalous. Until a short time ago our general 
knowledge of the CO distribution in galaxies was still based on the single-dish survey 
of 300 galaxies observed with the FCRAO 14-m Telescope (Young et al. 1995). 
They found that CO is usually peaked toward galaxy centres and only for the 
$\sim 10$\% 
of their systems there is an indication of a molecular ring or 
off-centre CO peak. In a more recent and higher resolution work, however, 
it appears that the picture is not so simple: the BIMA Survey of 
Nearby Galaxies (BIMA SONG), that studied the CO emission of 44 
nearby spiral galaxies, showed that the molecular gas distribution 
is generally very heterogeneous in this kind of systems (Regan et al. 
2001, Helfer et al. 2003). This survey suggests that spiral galaxies 
can show large departures from a smooth exponential profile
in the gaseous components. The possibility to find CO emission 
beyond the optical radius is also supported by the 
results obtained by Braine \& Herpin (2004) for the isolated spiral 
galaxy NGC4414.  The spectra that they measured reveal the presence of 
cool molecular gas out 1.5 times the optical radius. 
In this galaxy, the CO found at large radii
is compatible with an exponential profile.
However, we know today that gas disks may have irregular radial 
distributions, with sometimes holes in the center, 
contrasted spiral structure, resonant rings, etc.. 

\begin{figure*}[ht]
\centering
\begin{tabular}{l l l l}
  \includegraphics[height=0.47\textwidth,angle=-90]{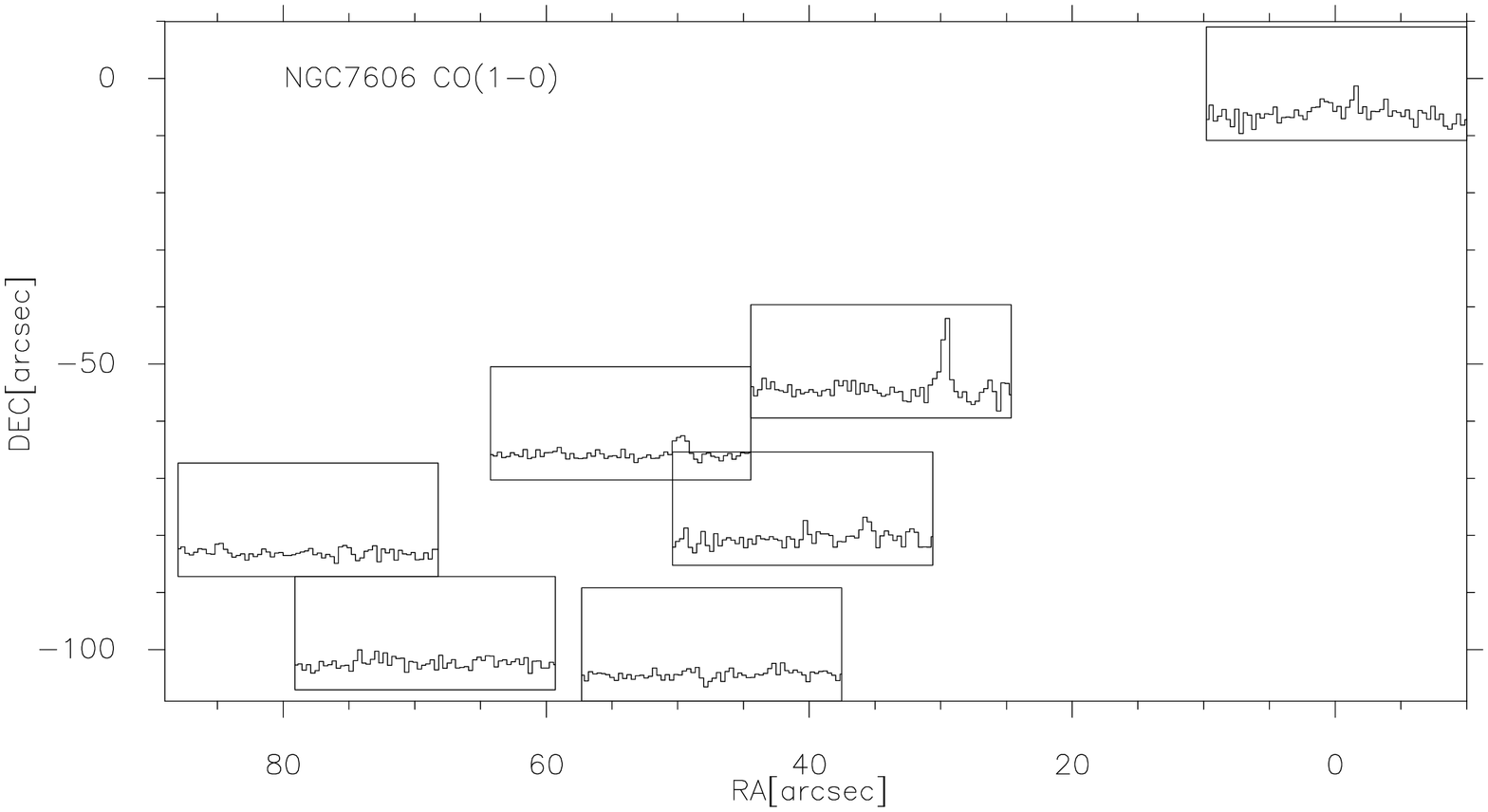}&
 \includegraphics[height=0.47\textwidth,angle=-90]{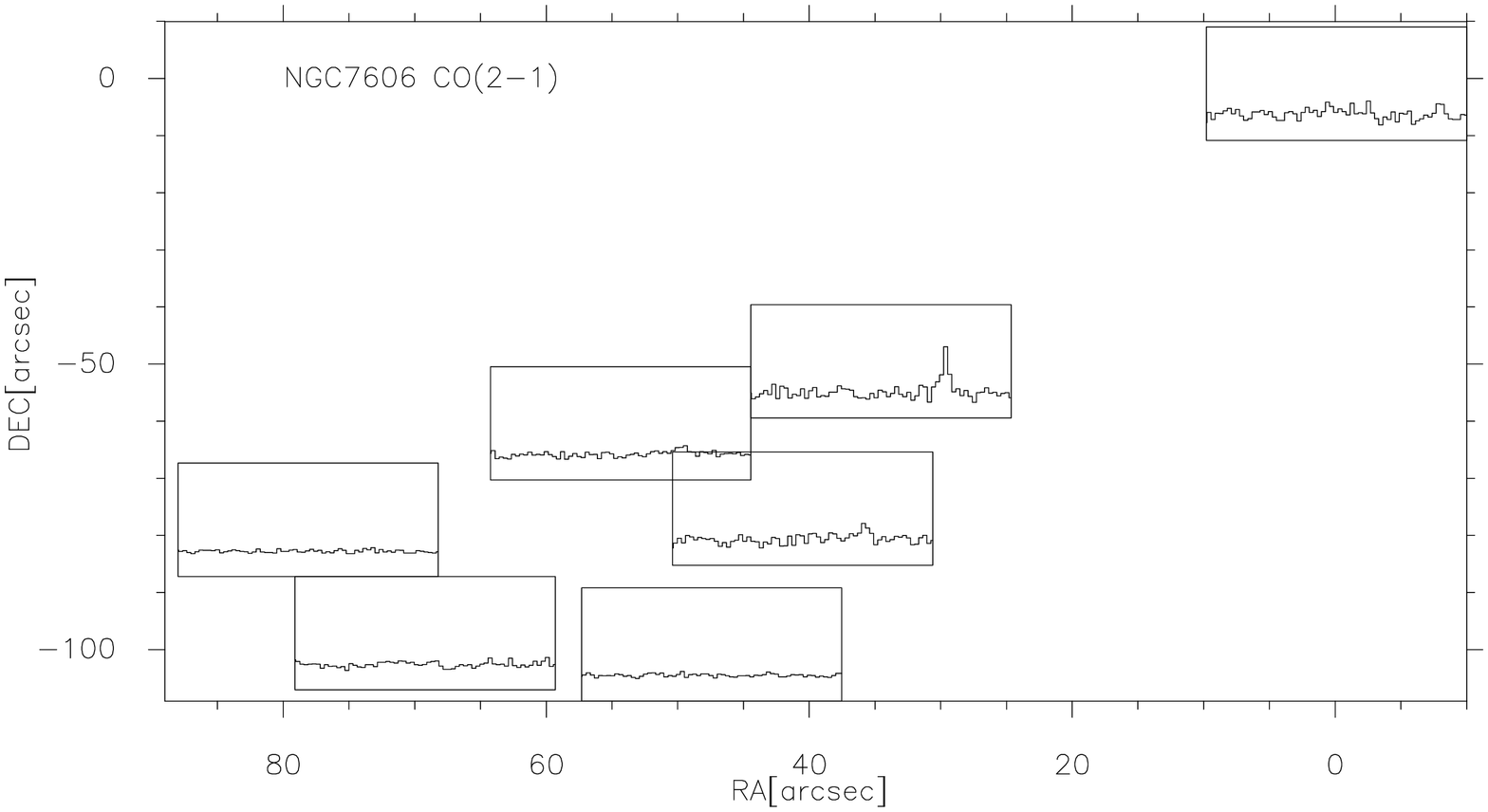}\\
\includegraphics[height=0.47\textwidth,angle=-90]{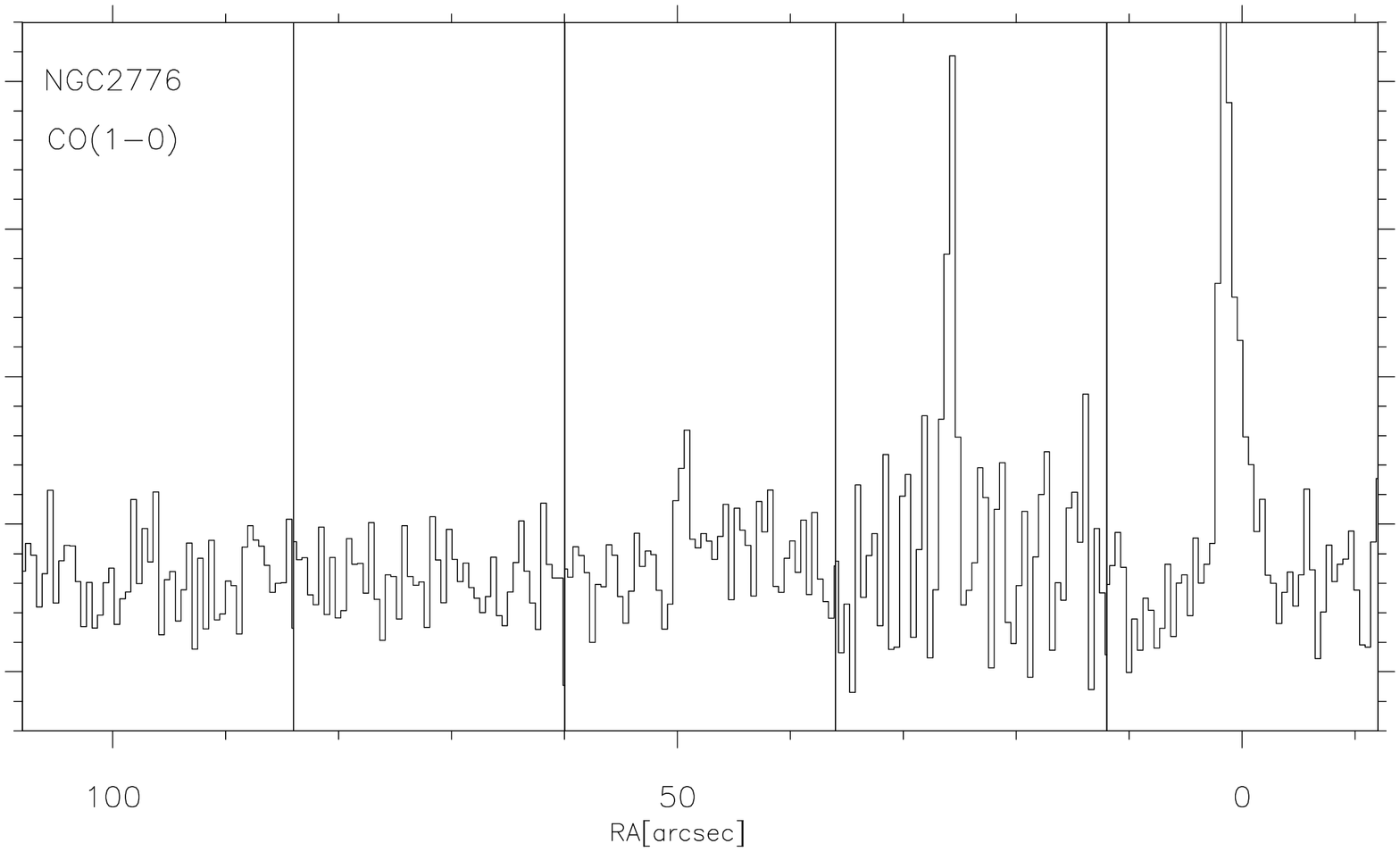}&
 \includegraphics[height=0.47\textwidth,angle=-90]{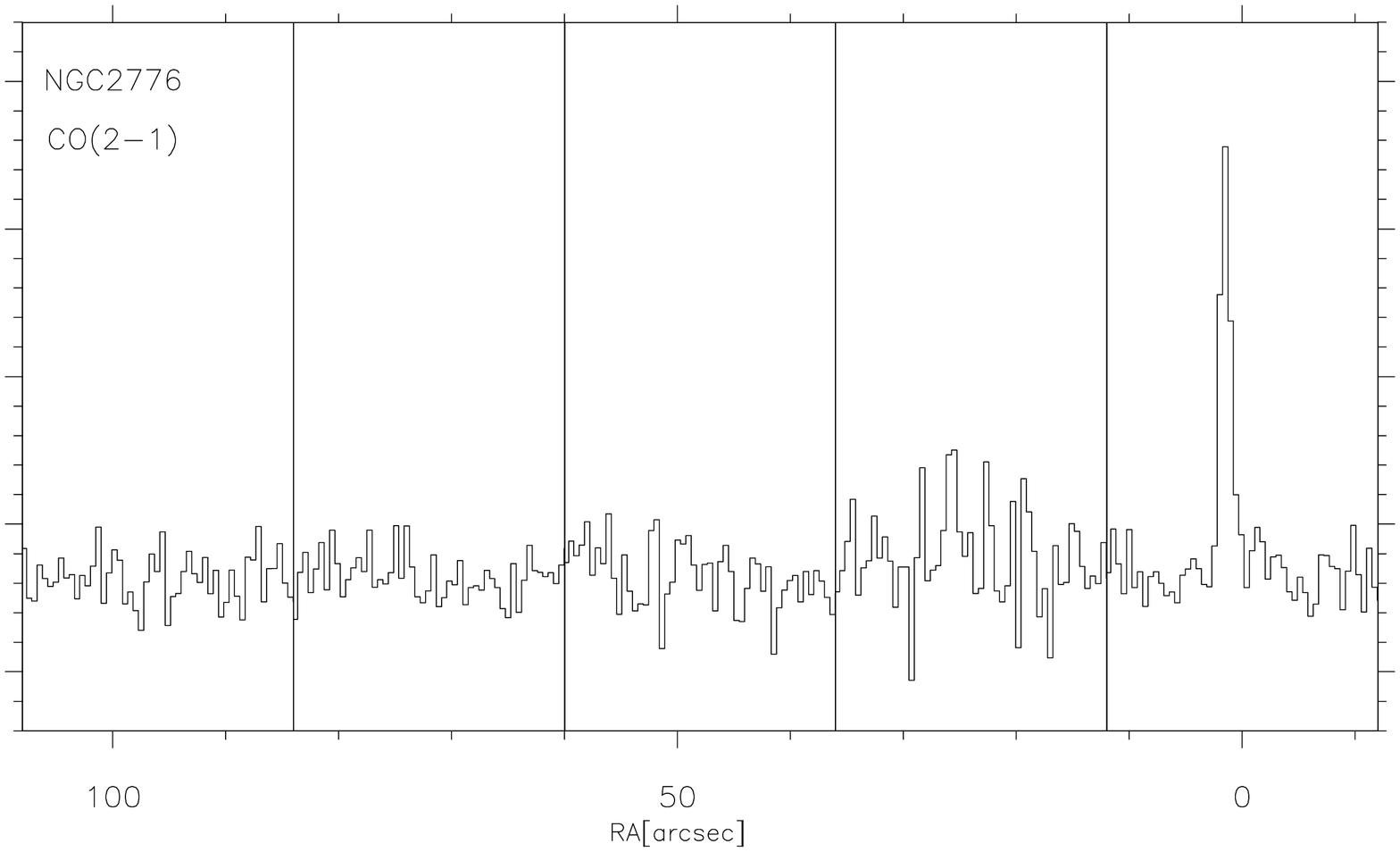}
\end{tabular}
\caption{First results obtained. Maps of spectra are plotted, RA vs. DEC 
in arcsec. Axes of each single spectrum: heliocentric radial velocity 
(v$_{h}$) in km/s (scale -500/500 km/s) and the main beam brightness temperature 
(T$_{mb}$) in K (scale -0.02/0.07 K). NGC7606: offsets (0,0), (35,-49), (41,-75), 
(55,-60), (79,-77), (70,-97), (48,-99). NGC27676: offsets (0,0), (24,0),
(48,0), (72,0), (96,0). {\it Top-left}: NGC7606 CO(1-0) maps.
The nucleus shows a weak emission, while the offset (35,-49) has a strong
signal. {\it Top-right}: NGC7606 CO(2-1) map. The results
are similar to those of the CO(1-0) line. {\it Bottom-left}: NGC2776
CO(1-0) map. Strong signal in the nucleus and in the offset
(24,0), then the signal decreases with the radius. {\it Bottom-right}:
NGC2776 CO(2-1) map. Similar results of CO(1-0) line but with
minor intensity.
}
\label{fig:2}       
\end{figure*}

\section{Conclusions}
\label{sec:4}
These first results encourage us to continue to explore with new observations 
the extension of CO beyond the optical radius and to see whether this is 
different for truncated or untruncated galaxies. This possible difference 
could allow us to test the theory that assumes the existence 
of a critical CO density acting as a star formation threshold. 
We will also compare with other gas tracers in the outer parts
of galaxies. Indeed the CO tracers might suffer from caveats:
the CO line emission requires excitation, that could be lacking
in the absence of new stars around. CO emission also depends on
metallicity, which is decrasing in the outer parts of galaxies.

\end{document}